# Electron acoustic shock and solitary waves in spin polarized dense rotating quantum plasmas

Atherv Saxena[1] and Punit Kumar[1*]

[1]*Department of Physics, University of Lucknow, India*

[1*]*E-mail- kumar_punit@lkouniv.ac.in*

## Abstract

The propagation of electrostatic waves in a three-component electron-positron-ion (e-p-i) astrophysical quantum plasma under the influence of uniform rotation is analyzed, incorporating the effects of particle spin, Fermi pressure, and the quantum Bohm potential. Spin polarization arising due to the alignment of particle spins under the influence of a strong external magnetic field, leads to an imbalance in the population of spin-up and spin-down states. Additionally, key astrophysical factors such as rotation and gravitational influence have been considered. The coupled dispersion relations for electron, positron, and ion modes have been derived. Further, the electron acoustic shock wave is studied using the Korteweg–de Vries–Burgers (KdVB) method, and the shock wave solution has been obtained. Quantum effects are found to contribute to enhance wave dispersion and modify the shock profile by broadening and stabilizing the shock structure.

## 1. Introduction

The study of electron-acoustic waves (EAWs) has increasingly expanded into the domain of quantum plasmas, where quantum effects play crucial roles. EAWs arise in plasmas containing two distinct electron temperatures and propagate at frequencies much higher than the ion plasma frequency, with phase velocities between the thermal velocities of cold and hot electron populations [1]. These two-electron-temperature conditions, common in non-equilibrium laboratory systems such as fusion devices, sputtering magnetron plasmas, and laser-plasma interactions, and in space environments like the solar wind, Earth's bow shock, and magnetospheres of planets and neutron stars [2, 3], are the cause for EAW excitation. While the existence of EAWs was initially confirmed in classical plasmas through laboratory experiments [4–6] and early theoretical models [7, 8], recent research has emphasized their propagation in quantum environments. Growing interest has focused on the nonlinear behaviour of EAWs in quantum plasmas, especially in interpreting structures observed in spacecraft missions such as FAST, GEOTAIL, and POLAR [9–22].

Quantum plasmas are found in a wide range of astrophysical environments such as white dwarfs [23], neutron stars [24], magnetars [25], and the early universe [26], where high densities and strong magnetic fields make quantum effects like Fermi degeneracy and spin polarization significant. In laboratory settings, quantum plasmas are studied in systems such as solid-state plasmas [27], ultra-cold plasmas [28], metallic nanostructures [29], intense laser-solid interactions [30], and advanced semiconductor devices [31], where quantum diffraction and collective behaviours are observable. A subset of these plasmas is electron-positron-ion (e–p–i) plasmas, where presence of positrons introduce unique dynamics due to their equal mass and opposite charge to electrons [32–36], differing fundamentally from conventional electron–ion (e–i) plasmas. Such pair plasmas, believed to have existed shortly after the Big Bang, significantly influence electrostatic and electromagnetic wave modes and are found in extreme astrophysical environments [37, 38] like polar caps of neutron stars, AGNs, solar flare regions [39], the Galactic center [40], and pulsar magnetospheres [41]. Their origins are linked to phenomena like intense laser-plasma interactions, white dwarf collapses, pulsar radio emissions, and quasar jets, as well as results from space missions such as PAMELA [42–48]. Observations confirm that ions often coexist with electrons and positrons, forming e–p–i plasmas [49], where ions substantially modify plasma dynamics. Despite high pair annihilation rates, positrons can persist under certain conditions, particularly in white dwarfs. Recent studies have explored various aspects of e–p and e–p–i plasmas [50–53], including the effects of ion temperature on large-amplitude ion-acoustic waves [54, 55], the behaviour of two-dimensional magnetosonic waves with varying positron concentrations [56], and the influence of quantum effects on ion-acoustic waves using KdV and KP equations [57, 58].

Quantum plasmas have attracted significant attention due to quantum mechanical effects, which become pronounced when the de Broglie wavelength of particles is comparable to the inter-particle distance [59–61]. The Quantum Hydrodynamic (QHD) model is widely used for its computational efficiency, simplifying boundary conditions by employing macroscopic variables [62, 63]. However, earlier QHD versions, which treated spin-up and spin-down particles identically, contradicted the Pauli Exclusion Principle [64]. The Separated Spin Evolution Quantum Hydrodynamic (SSE-QHD) model resolves this by considering spin-up and spin-down particles as distinct components [65], discovering new wave modes [66–68], and improving our understanding of magnetic field interactions in plasma waves [69,70]. This makes the SSE-QHD model a more accurate framework for analyzing quantum plasmas [71–73]. Building on this, the study of dense rotating plasmas, where spin polarization and rotation are key, becomes essential for exploring complex plasma behaviour in astrophysical contexts. Applying the quantum hydrodynamic (QHD) model, researchers have analyzed the dispersion properties and the formation of bright and dark electron acoustic solitons in dense, unmagnetized quantum plasmas with immobile ions and two electron species [74]. Planar geometry studies have revealed modulational instability in obliquely propagating EAWs, leading to both bright and dark envelope structures [75], while the dynamics of electron-acoustic solitary waves in magnetized quantum plasmas with two-temperature electron distributions have also been explored [76]. Moreover, simulation results involving quantum electron plasmas with fixed ion backgrounds have shown the existence of one-dimensional dark envelope solitons and quantum electron vortices stable at lower charge states and unstable at higher ones [77].

High-density plasmas under rotation and magnetic fields play a critical role in both astrophysical and experimental contexts. In compact stellar objects, strong rotation combined with intense magnetic fields is common, especially during stellar collapse, where angular momentum conservation leads to increased rotational speed and the generation of a rapidly rotating, highly magnetized plasma environment. Often, magnetic field and rotation axes intersect at an inclination angle [78–82]. Such configurations are also observed in geophysical settings [83] and advanced laboratory experiments, including tokamak devices [84–86]. While particle densities in geophysical and laboratory plasmas are lower than those in astrophysical systems, the effects of rotation, particularly the Coriolis force, are significant, emulating the influence of a magnetic field on the ionized medium [87–93].

This paper focuses on analyzing the coupled dispersion relations in multi-component plasmas and exploring the formation of solitary wave structures. Section 2 outlines the fundamental equations of the Separated Spin Evolution Quantum Hydrodynamic (SSE-QHD) model along with particle dynamics. In Section 3, the coupled dispersion characteristics of electrons, positrons, and ions in a rotating, magnetized e–p–i quantum plasma are examined using the SSE-QHD framework. Section 4 investigates the electron acoustic shock wave phenomena by employing the Korteweg–de Vries Burgers (KdVB) approach and deriving corresponding shock solutions. Lastly, Section 5 is devoted to summary and discussion.

## 2. Quantum Plasma Dynamics

We consider a collisionless electron-positron-ion quantum plasma composed of two distinct populations of electrons, low temperature, inertial cold electrons, and high temperature, inertia - less hot electrons, positrons and ions. Quantum plasma is embedded in a constant external magnetic field acting along the z direction $\left(\vec{B} = B_0(\hat{z})\right)$, and an external electrostatic wave $\vec{E} = -\nabla\phi$, where $\phi$ is the electrostatic potential interacts with the plasma. The plasma is taken to be rotating in the astrophysical settings with angular frequency $\vec{\Omega}$, at an angle $\theta$ to the direction of the magnetic field. The basic set of governing QHD fluid eqs. comprises of continuity eqn., momentum eqn., and the system is closed by the Poisson eqn. as given below,

$$\frac{\partial n_{j\alpha}}{\partial t} + \frac{\partial}{\partial x}\left(n_{j\alpha}\vec{v}_{j\alpha}\right) = 0,$$

(1a)

$$\frac{\partial \vec{v}_{j\alpha}}{\partial t} + \vec{v}_{j\alpha}\frac{\partial \vec{v}_{j\alpha}}{\partial x} = \frac{e}{m_j}\frac{\partial \phi}{\partial x} - \frac{e}{m_j}\left(\vec{v}_{j\alpha}\times\vec{B}\right) - \frac{1}{m_j n_{0j}}\frac{\partial}{\partial x}\vec{P}_{jf\alpha} + \frac{\hbar^2}{2m_j^2}\frac{\partial}{\partial x}\left(\frac{1}{\sqrt{n_{j\alpha}}}\frac{\partial^2}{\partial x^2}\sqrt{n_{j\alpha}}\right) + 2\left(\vec{v}_{j\alpha}\times\vec{\Omega}\right) - \frac{\partial}{\partial x}\Phi_j,$$

(1b)

$$\varepsilon_0 \frac{\partial^2 \phi}{\partial x^2} = \sum_{j=c,h,p,i} q_j n_{j\alpha}.$$

(1c)

where, $v_j$, $m_j$, $n_j$, $P_j$ represent the fluid velocity, rest mass, particle density and Fermi pressure of the $j^{th}$ species of plasma i.e. cold electrons $(ce)$, hot electrons $(he)$, positrons $(p)$, and ions $(i)$. $\alpha = \uparrow$ and $\downarrow$ denotes spin-up and spin-down fermions respectively. Eqn. (1b) corresponds to momentum eqn. for plasma species $j$. The second term, on the left hand side of eq. (1b) is the convective derivative of the velocity. The first term, on the right hand side of equation (1b) refers to the Lorentz force, the second term is the force due to the degenerate pressure $P_{j\alpha} = \left(m_j V_{Fj\alpha}^2 / 3n_{j0}^2\right) n_{j\alpha}^3$, where $V_{Fj\alpha} = \sqrt{\zeta_\alpha} V_{Fj}$ is the Fermi velocity of fermions with $V_{Fj} (= \hbar(3\pi^2 n_{0j})^{1/3} / m_j)$ and $\zeta_\alpha = \left\{(1-\eta)^{5/3} + (1+\eta)^{5/3}\right\}/2$, and $\eta (= \Delta n_{j\alpha}/n_0)$ is the spin polarization due to the presence of magnetic field, $\Delta n_{j\alpha} = \sum_j (n_\uparrow - n_\downarrow)$ denotes the concentration difference of spin-up $(\uparrow)$ and spin-down $(\downarrow)$ fermions. The third term is the quantum Bohm force involving quantum electron tunneling in dense quantum plasma. The fourth term, is the Coriolis force, due to the rotation of the plasma with angular velocity $\vec{\Omega}$. Since, rotation is taken to be slow, quadratic and higher order terms such as centrifugal force $\sim \vec{\Omega}\times(\vec{\Omega}\times\vec{r})$ are safely neglected. The last term is the gravitational potential term which is derived from the Poisson's equation for gravitational potential field as, $\nabla^2 \Phi_j = 4\pi G n_j$, where $G$ is the gravitational constant.

The cold electrons, due to their low temperature and reduced mobility, provide the inertial contribution to the dynamics, while the hot electrons, with their higher temperature and greater mobility, offer the restoring force. So, in the momentum equation for hot electrons, we consider inertia term as zero. The phase speed of the EAW lies in the range $v_{Fc} \ll \omega/k \ll v_{Fh}$, where $v_{Fc}$ and $v_{Fh}$ are the Fermi velocities of cold and hot electrons, respectively. Since $n_{0c} \ll n_{0h}$ holds for EAW, which implies that $T_{Fc} \ll T_{Fh}$ in quantum plasmas, therefore the Fermi pressure due to cold electrons can be ignored in comparison to the hot electrons in the model.

The equations describing the motion of plasma species are given as,

$$\frac{\partial \vec{v}_{ce\alpha}}{\partial t} + \vec{v}_{ce\alpha} \frac{\partial \vec{v}_{ce\alpha}}{\partial x} = \frac{e}{m_e}\left(\frac{\partial \phi}{\partial x} - \left(\vec{v}_{ce\alpha} \times \vec{B}\right)\right) + \frac{\hbar^2}{2m_e^2} \frac{\partial}{\partial x}\left(\frac{1}{\sqrt{n_{ce\alpha}}} \frac{\partial^2}{\partial x^2} \sqrt{n_{ce\alpha}}\right) + 2\left(\vec{v}_{ce\alpha} \times \vec{\Omega}\right), \tag{2}$$

$$0 = \frac{e}{m_e}\left(\frac{\partial \phi}{\partial x} - \left(\vec{v}_{he\alpha} \times \vec{B}\right)\right) - \frac{1}{m_e n_{he0}} \frac{\partial P_{Fh\alpha}}{\partial x} + \frac{\hbar^2}{2m_e^2} \frac{\partial}{\partial x}\left(\frac{1}{\sqrt{n_{he\alpha}}} \frac{\partial^2}{\partial x^2} \sqrt{n_{he\alpha}}\right) + 2\left(\vec{v}_{he\alpha} \times \vec{\Omega}\right), \tag{3}$$

$$0 = -\frac{e}{m_p}\left(\frac{\partial \phi}{\partial x} - \left(\vec{v}_{p\alpha} \times \vec{B}\right)\right) - \frac{1}{m_p n_{p0}} \frac{\partial P_{Fp\alpha}}{\partial x} + \frac{\hbar^2}{2m_p^2} \frac{\partial}{\partial x}\left(\frac{1}{\sqrt{n_{p\alpha}}} \frac{\partial^2}{\partial x^2} \sqrt{n_{p\alpha}}\right) + 2\left(\vec{v}_{p\alpha} \times \vec{\Omega}\right), \tag{4}$$

In the case of ions, due to their large mass as compared to the electrons and positrons, i.e., $m_{e,p}/m_i \ll 1$, the quantum effects are insignificant and so they can be considered classically as [54],

$$\frac{\partial \vec{v}_i}{\partial t} + \vec{v}_i \frac{\partial \vec{v}_i}{\partial x} = -\frac{e}{m_i}\left(\frac{\partial \phi}{\partial x} - (\vec{v}_i \times \vec{B})\right) + 2(\vec{v}_i \times \vec{\Omega}) - \frac{\partial}{\partial x}\Phi, \tag{5}$$

and the Poisson's eq. become,

$$\frac{\partial^2 \phi}{\partial x^2} = \frac{e}{\varepsilon_0}\left(n_{ce} + n_{he} - n_p - n_i\right). \tag{6}$$

Eqs. (2), (3), (4) and (5) correspond to momentum equation for cold electrons, hot electrons, positrons and ions respectively and eqn. (6) is the Poisson's equation.

## 3. Coupled e-p-i dispersion relation

Perturbatively expanding eqs. (1–6) in orders of the fields of the electrostatic wave and assuming all the varying parameters to take the form,

$$f = f_0 + \chi f^{(1)} + \chi^2 f^{(2)}. \tag{7}$$

with $f_0$ representing the initial value, $f^{(1)}$ is the first order perturbation term and $f^{(2)}$ is the second order perturbation and $\chi$ is the dimensionless parameter and assuming first order perturbed quantities to vary as $e^{i(kx-\omega t)}$, we get the corresponding first order perturbed velocities as,

$$\vec{v}_{ce\alpha x}^{(1)} = -\frac{e\omega k}{m_e(\omega^2 - \omega_{eff}^2)}\phi^{(1)} + \frac{Q_{ce\alpha}k^3\omega}{(\omega^2 - \omega_{eff}^2)}n_{ce\alpha x}^{(1)}, \tag{8}$$

$$\vec{v}_{ce\alpha y}^{(1)} = \frac{ie\omega_{eff}k}{m_e(\omega^2 - \omega_{eff}^2)}\phi^{(1)} - \frac{iQ_{ce\alpha}k^3\omega_{eff}}{(\omega^2 - \omega_{eff}^2)}n_{ce\alpha x}^{(1)}, \tag{9}$$

$$\vec{v}_{he\alpha}^{(1)} = \frac{eik}{m_e\omega_{eff}}\phi^{(1)} - \frac{Q_{he\alpha}ik^3}{\omega_{eff}}n_{he\alpha}^{(1)} - \frac{V_{Fh\alpha}^2 ik}{n_{he0}\omega_{eff}}n_{he\alpha}^{(1)}, \tag{10}$$

$$v_{p\alpha}^{(1)} = \frac{eik}{m_p \omega_{eff}} \phi^{(1)} + \frac{Q_{p\alpha} ik^3 n_{p\alpha}^{(1)}}{\omega_{eff}} + \frac{V_{Fp\alpha}^2 ik}{n_{p0} \omega_{eff}} n_{p\alpha}^{(1)}, \tag{11}$$

$$v_{ix}^{(1)} = \frac{e\omega}{m_i \left(\omega^2 + \omega_{eff}^2\right)} \phi^{(1)} + \frac{4\pi G ik^2 \omega}{\left(\omega^2 + \omega_{eff}^2\right)} n_{ix}^{(1)}, \tag{12}$$

$$v_{iy}^{(1)} = -\frac{ie\omega_{eff}}{m_i \left(\omega^2 + \omega_{eff}^2\right)} \phi^{(1)} + \frac{4\pi G k^2 \omega_{eff}}{\left(\omega^2 + \omega_{eff}^2\right)} n_{ix}^{(1)}, \tag{13}$$

where, $Q_j \left(= \frac{\hbar^2}{4m_j^2 n_{j0}}\right)$ and $n_{j0}$ is the equilibrium number density of different plasma species, and $\omega_{eff} = \left(\omega_{cj} + 2\Omega_0 \cos\theta\right)$ is the effective angular frequency due to rotation, in terms of cyclotron frequency of plasma species $\omega_{cj} = eB_0/m_j$.

Simultaneously solving eqs. (8-13), we arrive at the following dispersion relation of the EAW in the multispecies magnetized astrophysical quantum plasma,

$$k^2 = \frac{ek^2 \omega_{pce}}{\omega^2 - \omega_{eff}^2 - Q_{ce\alpha} k^4 n_{ce0}} - \frac{iek^2 \omega_{phe}}{\omega\omega_{eff} + Q_{he\alpha} n_{h0} ik^4 + V_{Fh\alpha}^2 ik^2} + \frac{iek^2 \omega_{pp}}{\omega\omega_{eff} - Q_{p\alpha} n_{p0} ik^4 - V_{Fp\alpha}^2 ik^2} + \frac{ek^2 \omega_{pi}}{\omega^2 + \omega_{eff}^2 - 4\pi G i n_{i0} k^3}$$

(14)

Equation (14) represents the dispersion relation for electron acoustic waves (EAWs) in a multispecies magnetized astrophysical quantum plasma. This shows, how the different species (cold electrons, hot electrons, positrons and ions) contribute to the propagation of electron acoustic waves in plasma. On the right-hand side, the first term, is the contribution of cold electrons. The second term, accounts for the influence of hot electrons. The third term, describes the contribution of positrons. The fourth term, signifies the role of ions in the plasma where, $\omega_{pj}$ is the plasma frequency of $j^{th}$ plasma species. Each term contributes uniquely to the overall dispersion relation, reflecting the interplay between different plasma species and the quantum effects that govern their dynamics. Numerical investigations to follow have adopted astrophysical plasma conditions typical of the outer layers of compact stars like white dwarfs or neutron stars. These parameters include ion number density $10^{26} m^{-3} \leq n_{0i} \leq 10^{28} m^{-3}$, the cold electron number density $n_{ce0} = 0.375 \times 10^{28} m^{-3}$ and the hot electron number density $n_{he0} = 0.125 \times 10^{28} m^{-3}$ where pair annihilation effects are negligible in such dense electron-positron plasmas, magnetic field strength $B_0 = 10 - 10^5 T$, and $T_{Fe} \approx 10^7 - 10^8 K$ [35-40].

Fig. 1 shows the variation of normalized propagation vector $kc/\omega_{pce}$ with normalized wave frequency $\omega/\omega_{pce}$. Normalized propagation vector reflects the direction and magnitude of wave propagation, while normalized wave frequency indicates wave transmissibility within the plasma medium. The quantum dispersion curve exhibits a significantly steeper increase in frequency with increasing wave number compared to absence of quantum effects. This behaviour highlights the enhanced wave transmissibility and phase velocity introduced by quantum mechanical effects. Specifically, the Bohm potential contributes a quantum diffraction term that becomes increasingly significant at shorter wavelengths (higher $k$), while the Fermi pressure accounts for degeneracy effects associated with dense electron and positron populations.

Fig. 2 shows the variation of normalized propagation vector $kc/\omega_{pce}$ with normalized wave frequency $\omega/\omega_{pce}$ for different values of spin polarization parameter $(\eta)$. The graph shows that increasing $\eta$ leads to a steeper dispersion curve, indicating that for a given wave frequency, the wavenumber increases with spin polarization. The solid curve corresponds to strong spin polarization ($\eta=0.8$), where these effects are dominant, resulting in a pronounced rise in wave number and thus a higher degree of wave dispersion and localization. The dashed line, representing a weakly polarized plasma ($\eta=0.008$), shows moderate dispersion effects, while the black dotted curve for $\eta=0$ (unpolarized plasma) exhibits the slowest growth in wave number, indicating minimal spin contribution. This trend implies that spin polarization not only alters the phase velocity of waves, but also increases the plasma's effective rigidity, making wave propagation more dispersive.

Fig. 3 shows the variation of normalized propagation vector $kc/\omega_{pce}$ with normalized wave frequency $\omega/\omega_{pce}$ for different values of cold to hot electron population $(n_{ce0}/n_{he0})$. As the ratio increases, the cold electron population becomes more dominant relative to hot electrons. This trend results in a reduction in the wavenumber $k$ for a given frequency $\omega$, i.e., the dispersion curves shift downward. Cold electrons, being inertial and more localized, provide the necessary inertia for electron acoustic wave propagation. Hot electrons are treated as inertialess and supply the restoring force via their thermal pressure. An increased cold electron population enhances the inertia in the system, thereby lowering the wave number required for a given frequency. This leads to reduced phase speed and less dispersion. Conversely, when $(n_{ce0}/n_{he0})$ is small, the hot electron pressure dominates, allowing the wave to propagate more easily with higher wavenumber values at the same frequency, indicating greater transmissibility and sharper dispersion.

## 4. Electron acoustic shock wave (EASW)

In order to study the nonlinear behaviour of electron acoustic wave, we use standard perturbation technique and introduce the stretched coordinates,

$$\zeta = \chi^{1/2}(x - \lambda t), \tag{15}$$

and

$$\tau = \chi^{3/2} t, \tag{16}$$

where, $\lambda$ is the phase velocity of wave. In this transformation, $x$ and $t$ are function of $\zeta$ and $\tau$ respectively, so partial derivatives with respect to $x$ and $t$ can be transformed into partial derivative in terms of $\zeta$ and $\tau$ as,

$$\frac{\partial}{\partial x} = \chi^{1/2} \frac{\partial}{\partial \zeta}, \tag{17}$$

$$\frac{\partial}{\partial t} = -\chi^{1/2} \lambda \frac{\partial}{\partial \zeta} + \chi^{3/2} \frac{\partial}{\partial \tau}, \tag{18}$$

$$\frac{\partial^2}{\partial x^2} = \chi \frac{\partial^2}{\partial \zeta^2}, \tag{19}$$

and

$$\frac{\partial^3}{\partial x^3} = \chi^{3/2} \frac{\partial^3}{\partial \zeta^3}. \tag{20}$$

We can express eqs. (1-6) in terms of $\zeta$ and $\tau$ as,

$$0 = -\chi^{1/2} \lambda \frac{\partial n_{j\alpha}}{\partial \zeta} + \chi^{3/2} \frac{\partial n_{j\alpha}}{\partial \tau} + \chi^{1/2} \frac{\partial}{\partial \zeta}\left(n_{j\alpha} \vec{v}_{j\alpha}\right) \tag{21}$$

$$-\lambda \chi^{1/2} \frac{\partial \vec{v}_{ce\alpha}}{\partial \zeta} + \chi^{3/2} \frac{\partial \vec{v}_{ce\alpha}}{\partial \tau} + \chi^{1/2} \vec{v}_{ce\alpha} \frac{\partial \vec{v}_{ce\alpha}}{\partial \zeta} = \frac{e}{m_e}\left(\chi^{1/2} \frac{\partial \phi}{\partial \zeta} - \left(\vec{v}_{ce\alpha} \times \vec{B}\right)\right) + Q_{ce\alpha} \chi^{3/2} \frac{\partial^3 n_{ce\alpha}}{\partial \zeta^3} + 2\left(\vec{v}_{ce\alpha} \times \vec{\Omega}\right), \tag{22}$$

$$0 = \frac{e}{m_e}\left(\chi^{1/2} \frac{\partial \phi}{\partial \zeta} - \left(\vec{v}_{he\alpha} \times \vec{B}\right)\right) + Q_{he\alpha} \chi^{3/2} \frac{\partial^3 n_{he\alpha}}{\partial \chi^3} - \frac{V_{Fh\alpha}}{n_{he0}^2} n_{he\alpha} \chi^{1/2} \frac{\partial n_{he\alpha}}{\partial \zeta} + 2\left(\vec{v}_{he\alpha} \times \vec{\Omega}\right), \tag{23}$$

$$0 = -\frac{e}{m_p}\left(\chi^{1/2} \frac{\partial \phi}{\partial \zeta} - \left(\vec{v}_{p\alpha} \times \vec{B}\right)\right) + Q_{p\alpha} \chi^{3/2} \frac{\partial^3 n_{p\alpha}}{\partial \chi^3} - \frac{V_{Fp\alpha}}{n_{p0}^2} n_{p\alpha} \chi^{1/2} \frac{\partial n_{p\alpha}}{\partial \zeta} + 2\left(\vec{v}_{p\alpha} \times \vec{\Omega}\right), \tag{24}$$

$$-\lambda \chi^{1/2} \frac{\partial \vec{v}_i}{\partial \zeta} + \chi^{3/2} \frac{\partial \vec{v}_i}{\partial \tau} + \chi^{1/2} \vec{v}_i \frac{\partial \vec{v}_i}{\partial \zeta} = -\frac{e}{m_i}\left(\chi^{1/2} \frac{\partial \phi}{\partial \zeta} - \left(\vec{v}_i \times \vec{B}\right)\right) - \chi^{1/2} \frac{\partial \Phi}{\partial \zeta} + 2\left(\vec{v}_i \times \vec{\Omega}\right), \tag{25}$$

and

$$\chi \frac{\partial^2 \phi}{\partial \zeta^2} = \frac{e}{\varepsilon_0}\left(n_{ce\alpha} + n_{he\alpha} - n_{p\alpha} - n_i\right). \tag{26}$$

Substituting the perturbations defined by eqs. (7) in eqs. (21) – (26), and collecting the lowest order terms, we get

$$n_{j0} \frac{\partial v_{j\alpha}^{(1)}}{\partial \zeta} - \lambda \frac{\partial n_{j\alpha}^{(1)}}{\partial \zeta} = 0, \tag{27}$$

$$\frac{e}{m_e}\frac{\partial \phi^{(1)}}{\partial \zeta}+\lambda \frac{\partial v_{ce\alpha}^{(1)}}{\partial \zeta}=0, \tag{28}$$

$$\frac{e}{m_e}\frac{\partial \phi^{(1)}}{\partial \zeta}-\frac{v_{Fhe}^2}{n_{he0}}\frac{\partial n_{he}^{(1)}}{\partial \zeta}=0, \tag{29}$$

$$\frac{e}{m_p}\frac{\partial \phi^{(1)}}{\partial \zeta}+\frac{v_{Fp}^2}{n_{p0}}\frac{\partial n_p^{(1)}}{\partial \zeta}=0, \tag{30}$$

$$\frac{e}{m_i}\frac{\partial \phi^{(1)}}{\partial \zeta}-\lambda \frac{\partial v_i^{(1)}}{\partial \zeta}=0, \tag{31}$$

and

$$\frac{e}{\varepsilon_0}\left(n_{ce\alpha}^{(1)}+n_{he\alpha}^{(1)}-n_{p\alpha}^{(1)}-n_i^{(1)}\right)=0. \tag{32}$$

Integrating and simplifying the above equations, yield

$$n_{ce\alpha}^{(1)}=-\frac{en_{ce0}}{m_e\lambda^2}\phi^{(1)}, \tag{33}$$

$$n_{he\alpha}^{(1)}=\frac{en_{he0}}{m_e V_{Fh\alpha}^2}\phi^{(1)}, \tag{34}$$

$$n_{p\alpha}^{(1)}=\frac{-en_{p0}}{m_p V_{Fp\alpha}^2}\phi^{(1)}, \tag{35}$$

$$n_i^{(1)}=\frac{en_{i0}}{m_i\lambda^2}\phi^{(1)}, \tag{36}$$

and substituting eqs. (33-36) into Poisson's eqn. (32) the phase velocity of EAW is obtained as,

$$\lambda=\left(\frac{V_{Fh\alpha}^2 V_{Fp\alpha}^2\left(n_{ce0}-n_{i0}\right)}{m_i\left(n_{he0}-n_{p0}\right)}\right)^{1/2}. \tag{37}$$

Now substituting the perturbations defined by eqs. (7) in eqs. (21) – (26), and collecting the higher order terms, we get

$$-\lambda\frac{\partial n_{j\alpha}^{(2)}}{\partial \zeta}+\frac{\partial n_{j\alpha}^{(1)}}{\partial \tau}+n_{j0}\frac{\partial v_{j\alpha}^{(2)}}{\partial \zeta}+\frac{\partial}{\partial \zeta}\left(n_{j\alpha}^{(1)}v_{j\alpha}^{(1)}\right)=0, \tag{38}$$

$$-\lambda\frac{\partial v_{ce\alpha}^{(2)}}{\partial \zeta}+\frac{\partial v_{ce\alpha}^{(1)}}{\partial \tau}+v_{ce\alpha}^{(1)}\frac{\partial v_{ce\alpha}^{(1)}}{\partial \zeta}=\frac{e}{m_e}\frac{\partial \phi^{(2)}}{\partial \zeta}+Q_{ce\alpha}\frac{\partial^3 n_{ce\alpha}^{(1)}}{\partial \zeta^3}, \tag{39}$$

$$\frac{e}{m_e}\frac{\partial \phi^{(2)}}{\partial \zeta}+Q_{he\alpha}\frac{\partial^3 n_{he\alpha}^{(1)}}{\partial \zeta^3}-\frac{V_{Fh\alpha}^2}{n_{h0}}\frac{\partial n_{he\alpha}^{(2)}}{\partial \zeta}-\frac{V_{Fh\alpha}^2}{n_{he0}^2}\left(n_{he\alpha}^{(1)}\frac{\partial n_{he\alpha}^{(1)}}{\partial \zeta}\right)=0, \quad (40)$$

$$\frac{-e}{m_p}\frac{\partial \phi^{(2)}}{\partial \zeta}+Q_{p\alpha}\frac{\partial^3 n_{p\alpha}^{(1)}}{\partial \zeta^3}-\frac{V_{Fp\alpha}^2}{n_{p0}}\frac{\partial n_{p\alpha}^{(2)}}{\partial \zeta}-\frac{V_{Fp\alpha}^2}{n_{p0}^2}\left(n_{p\alpha}^{(1)}\frac{\partial n_{p\alpha}^{(1)}}{\partial \zeta}\right)=0, \quad (41)$$

$$-\lambda\frac{\partial v_i^{(2)}}{\partial \zeta}-\frac{\partial v_i^{(1)}}{\partial \tau}+v_i^{(1)}\frac{\partial v_i^{(1)}}{\partial \zeta}=-\frac{e}{m_i}\frac{\partial \phi^{(2)}}{\partial \zeta}-\frac{\partial \Phi^{(2)}}{\partial \zeta}, \quad (42)$$

and

$$\frac{\partial^2 \phi^{(1)}}{\partial \zeta^2}=\frac{e}{\varepsilon_0}\left(n_{ce\alpha}^{(2)}+n_{he\alpha}^{(2)}-n_{p\alpha}^{(2)}-n_i^{(2)}\right). \quad (43)$$

Solving eqs. (38) – (43) we finally obtain,

$$\frac{\partial \phi^{(1)}}{\partial \tau}+C_1\phi^{(1)}\frac{\partial \phi^{(1)}}{\partial \zeta}+C_2\frac{\partial^3 \phi^{(1)}}{\partial \zeta^3}=C_3\frac{\partial^2 \phi^{(1)}}{\partial \zeta^2}. \quad (44)$$

Eq. (44) is the required KdVB equation, where the nonlinearity coefficient $C_1$, the dispersive coefficient $C_2$ and the dissipative coefficient $C_3$ are,

$$C_1=\left(\frac{\dfrac{2e\omega_{pce}^2}{m_e\lambda}-\dfrac{e\omega_{phe}^2\lambda^3}{m_e V_{Fh\alpha}^4}+\dfrac{e\omega_{pp}^2\lambda^3}{m_e V_{Fp\alpha}^4}-\dfrac{e\omega_{pi}^2}{m_i\lambda}}{2\omega_{pi}^2-2\omega_{pce}^2}\right),$$

$$C_2=\left(\frac{\dfrac{\omega_{pce}^2 Q_{ce\alpha}n_{ce0}}{\lambda}+\dfrac{\omega_{phe}^2 Q_{he\alpha}n_{he0}\lambda^3}{V_{Fh\alpha}^2}+\dfrac{\omega_{pp}^2 Q_{p\alpha}n_{p0}\lambda^3}{V_{Fp\alpha}^2}-1}{2\omega_{pi}^2-2\omega_{pce}^2}\right),$$

and

$$C_3=\left(\frac{(\Omega_0 \cos\theta)^2}{\lambda}+\frac{4\pi G}{\lambda}\right).$$

To obtain a shock wave solution to the KdVB equation (44), we transform the stretched coordinates $\zeta$ and $\tau$ into one coordinate $\eta=\zeta-U\tau$, where $U$ is the speed of the shock wave. Boundary conditions are $\phi^{(1)}\to 0, \frac{\partial \phi^{(1)}}{\partial \eta}\to 0, \frac{\partial^2 \phi^{(1)}}{\partial \eta^2}\to 0$ in the unperturbed region $\eta\to\pm\infty$.

Hence, we drop the superscript $(1)$ for ease of notation. Thus, the shock solution to eq. (44) is the standard result,

$$\phi = \phi_0 \tanh\left(\frac{\eta}{\Delta}\right),$$

where, the shock amplitude $\phi_0$ and the shock width $\Delta$ are,

$$\phi_0 = \frac{2U}{C_1},$$

and

$$\Delta = \frac{2C_2}{C_3}.$$

When dissipation is strong $(C_3 \gg 0)$, shock waves form due to energy loss. As $C_3 \to 0$, energy is conserved, leading to the formation of solitons that maintain their shape over long distances and the KdVB equation will reduce to KdV equation,

$$\frac{\partial \phi^{(1)}}{\partial \tau} + C_1 \phi^{(1)} \frac{\partial \phi^{(1)}}{\partial \zeta} + C_2 \frac{\partial^3 \phi^{(1)}}{\partial \zeta^3} = 0, \tag{45}$$

and the standard soliton solution is,

$$\phi = \phi_0 \operatorname{sech}^2\left(\frac{\eta}{\sigma}\right),$$

where the soliton width $(\sigma)$ is,

$$\sigma = \sqrt{\frac{4C_2}{U}}.$$

In the limit where the dissipative term in the KdVB equation vanishes, the system reduces to the integrable KdV equation, showing a transition from a dissipative to a conservative dynamical regime. This implies that the quantum plasma supports solitary wave structures formed through a balance between nonlinear steepening and quantum dispersive effects, with no energy loss, thereby enabling stable propagation of soliton.

Fig. 4 illustrates the temporal evolution of the electrostatic potential $\phi(\zeta,\tau)$ associated with electron acoustic shock waves propagating through spin-polarized, magnetized electron-positron-ion (e-p-i) quantum plasma. As the scaled time parameter $\tau$ increases from 1 to 4, the shock profile becomes increasingly steep and sharp, signifying the nonlinear steepening of the wavefront, a characteristic behaviour of dispersive shock waves. This steepening reflects the dominance of nonlinear convective terms over dispersive quantum corrections, leading to enhanced wave distortion and potential localization. The presence of spin polarization modifies the effective inertia and pressure of the spin-up and spin-down populations, altering the dynamics of the cold electrons that form the inertial component of the wave. Further, the rotation introduces anisotropy in wave propagation and modifies the effective pressure and transport properties, while gravity contributes an additional restoring force that influences the potential structure. The observed increase in amplitude and sharpness with time suggests energy accumulation and the formation of strong electrostatic shocks.

Fig. 5 shows the variation between electrostatic potential $\phi$ and space variable $\zeta$ in the limit when the dissipative term from the KdVB equation $C_3 \rightarrow 0$, effectively eliminating dissipative effects such as collisions, thereby allowing the nonlinear and dispersive effects to balance perfectly resulting in the formation of a soliton. The figure represents the soliton profiles for both presence of the quantum (solid curve) and absence of quantum effects (dashed curve) regimes. The quantum soliton exhibits a higher peak amplitude and narrower width, indicating stronger localization of the electrostatic potential due to quantum effects like the Bohm potential and Fermi pressure. These quantum corrections introduce additional dispersion which, when balanced with nonlinearity, leads to sharper, more confined solitary structures compared to the broader and lower-amplitude classical soliton. Spin polarization further modifies the effective pressure and mass distribution of the cold electron fluid, which contributes to the soliton dynamics.

Fig. 6 is a 3D representation of shock profile in quantum plasma with the variation of electrostatic potential ($\phi$) and the stretched coordinates $\zeta$ and $\tau$. The yellow region at the crest of the profile highlights the sharp peak of the electrostatic potential, signifying the most intense localized energy zone. The surrounding green and blue regions represent the broadened spatial extent of the shock front, indicating the asymmetric evolution of the profile due to dissipative effects. The dark blue zones toward the edges mark the initiation and dissipation phases of the shock, revealing how the amplitude stabilizes and disperses over the considered time frame.

Fig. 7 is a 3D representation of soliton profile in quantum plasma with the variation of electrostatic potential ($\phi$) and the stretched coordinates $\zeta$ and $\tau$. A contour of the 3D structure is also mentioned describing the stability of soliton's peak for approximately 2 seconds and soliton structure for approximately 10 seconds, in the presence of coriolis force and gravitational effects. The yellow colour shows the stability of peak of soliton and the green and blue colours are representing the width of soliton and the dark blue colour shows starting and ending phase of soliton structure.

## 5. Summary and discussion

The dynamics of a uniform, magnetized astrophysical quantum plasma composed of electrons, positrons, and ions are analyzed with a focus on electron acoustic shock waves. The Separated Spin Evolution Quantum Hydrodynamic (SSE-QHD) model serves as the theoretical framework, accounting for quantum diffraction, quantum statistical pressure, and distinct spin-up and spin-down contributions of fermions. A generalized dispersion relation is derived for the coupled plasma species, incorporating various quantum effects and external conditions. The formation and evolution of electron acoustic shock structures are explored using the Korteweg–de Vries Burgers (KdVB) equation and its shock solution. Further, it is shown that in the limit where the dissipative term in KdVB equation approaches zero, the KdVB equation reduces to the standard KdV equation, which yields soliton solutions. Through a combination of analytical derivation and graphical investigation, the study aims to deepen the understanding of nonlinear wave propagation in quantum plasma environments.

Electron acoustic wave transmissibility increases significantly in quantum plasma due to the combined influence of quantum corrections and spin polarization. A comparative analysis indicates that the presence of Bohm potential and Fermi pressure enhances dispersion steepness by approximately 47% relative to the classical case at the normalised angular frequency of 8.5, confirming stronger wave localization and higher phase velocity. Increasing the spin polarization parameter $\eta$ from 0 to 0.8 results in a 53% rise in normalized wave number at a normalised frequency of 8.5, indicating elevated dispersion and localization due to spin-induced modifications in effective pressure. Further, increasing the cold-to-hot electron density ratio results in a 38% decrease in normalized wave number for a normalised frequency of 8.5, attributed to greater inertial dominance of cold electrons over hot thermal pressure. In the nonlinear regime, shock wave profiles show a steepening of the electrostatic potential by over 55% with increasing scaled time, reflecting energy accumulation and wavefront sharpening. Suppression of dissipative effects reduces the KdVB system to a pure KdV regime, where the soliton solution displays 61% higher amplitude and 33% narrower width in the quantum case, highlighting enhanced localization due to quantum dispersion and degeneracy pressure. Spin polarization further amplifies soliton amplitude and compresses the structure, indicating stronger confinement of electrostatic energy.

This study provides valuable insights into the behaviour of electron-positron plasmas under extreme astrophysical conditions, including those found in quasars, gamma-ray bursts, active galactic nuclei, pulsar magnetospheres, black hole accretion disks, white dwarf atmospheres, and the Van Allen radiation belts, as well as in the early universe and the galactic center. The developed theoretical framework enhances our understanding of the underlying physical mechanisms in such environments and aids in interpreting observational data from high-energy astrophysical sources, thereby contributing to a deeper comprehension of some of the universe's most energetic and enigmatic phenomena.

## Acknowledgement

Financial support from SERB – DST under MATRICS is gratefully acknowledged (grant no. : MTR/2021/000471).

**Figure Captions**

**Figure 1:** Variation of $\omega/\omega_{pce}$ with $kc/\omega_{pce}$ for coupled e-p-i modes with $h = 1.054 \times 10^{-34} J-s$ (solid line) with $\theta = 4°$, $\Omega_0 = 0.0003$, $B_0 = 2 \times 10^4 T$ and $h \to 0$ (dashed line) with thermal velocity $V_{Te} = 10^7 cm/\sec$.

**Figure 2:** Variation of $\omega/\omega_{pce}$ with $kc/\omega_{pce}$ for coupled e-p-i modes with $\theta = 4°$, $\Omega_0 = 0.0003$, $B_0 = 2 \times 10^4 T$, $\eta = 0.8$ (solid line), $\eta = 0.008$ (dashed line), $\eta = 0$ (dotted line).

**Figure 3:** Variation of $\omega/\omega_{pce}$ with $kc/\omega_{pce}$ for coupled e-p-i modes with $\theta = 4°$, $\Omega_0 = 0.0003$, $B_0 = 2 \times 10 T$ with $n_{ce0}/n_{he0} = 0.1$ (solid line), $B_0 = 2 \times 10^2 T$ with $n_{ce0}/n_{he0} = 0.2$ (dashed line) and $B_0 = 2 \times 10^4 T$ with $n_{ce0}/n_{he0} = 0.3$ (dotted line).

**Figure 4:** Variation of $\phi$ with $\zeta$ for shock solution l =0.2, k =0.8, $\theta = 4°$, $\Omega_0 = 0.0003$, $B_0 = 2 \times 10^4 T$ and $\tau = 1, 2, 3, 4$.

**Figure 5:** Variation of $\phi$ with $\tau$ for soliton solution l =0.2, k =0.8, $h = 1.054 \times 10^{-34} J-s$ (solid line), $\theta = 4°$, $\Omega_0 = 0.0003$, $B_0 = 2 \times 10^4 T$ and $h \to 0$ (dashed line)

**Figure 6:** Variation of $\phi$ with $\xi$ and $\tau$ for shock solution with, l =0.2, k =0.8, $\theta = 4°$, $\Omega_0 = 0.0003$, $B_0 = 2 \times 10^4 T$ in a 3D representation.

**Figure 7:** Variation of $\phi$ with $\xi$ and $\tau$ for soliton solution with, l =0.2, k =0.8, $\theta = 4°$, $\Omega_0 = 0.0003$, $B_0 = 2 \times 10^4 T$ in a 3D representation with its contour.

**Figures**

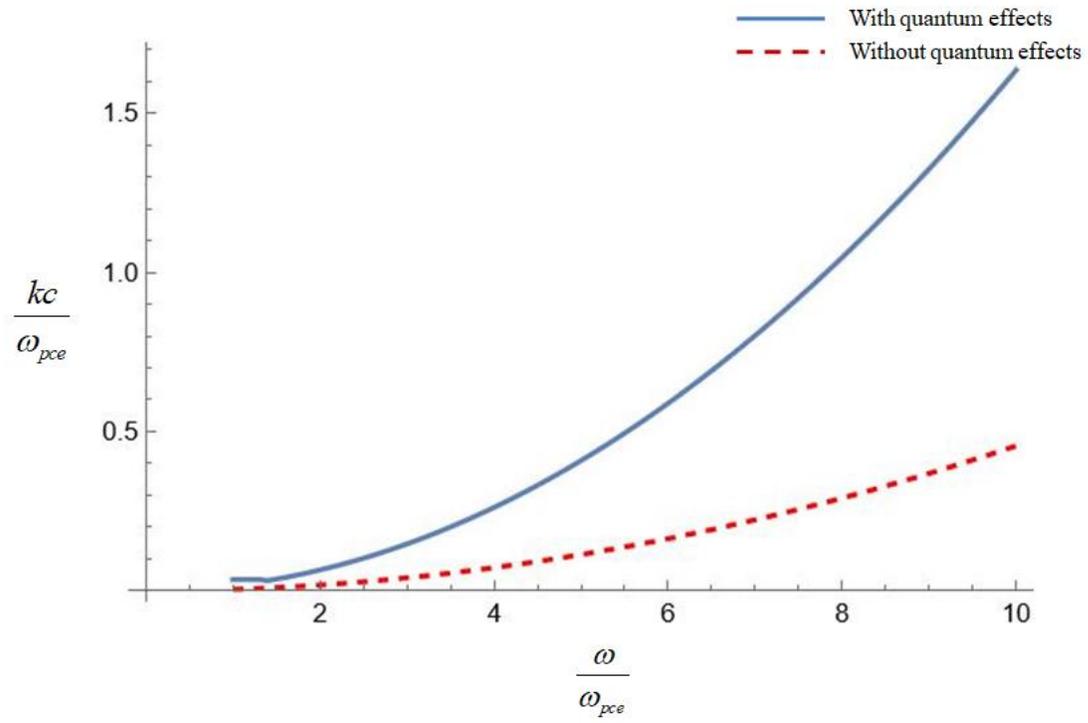

**Figure 1.**

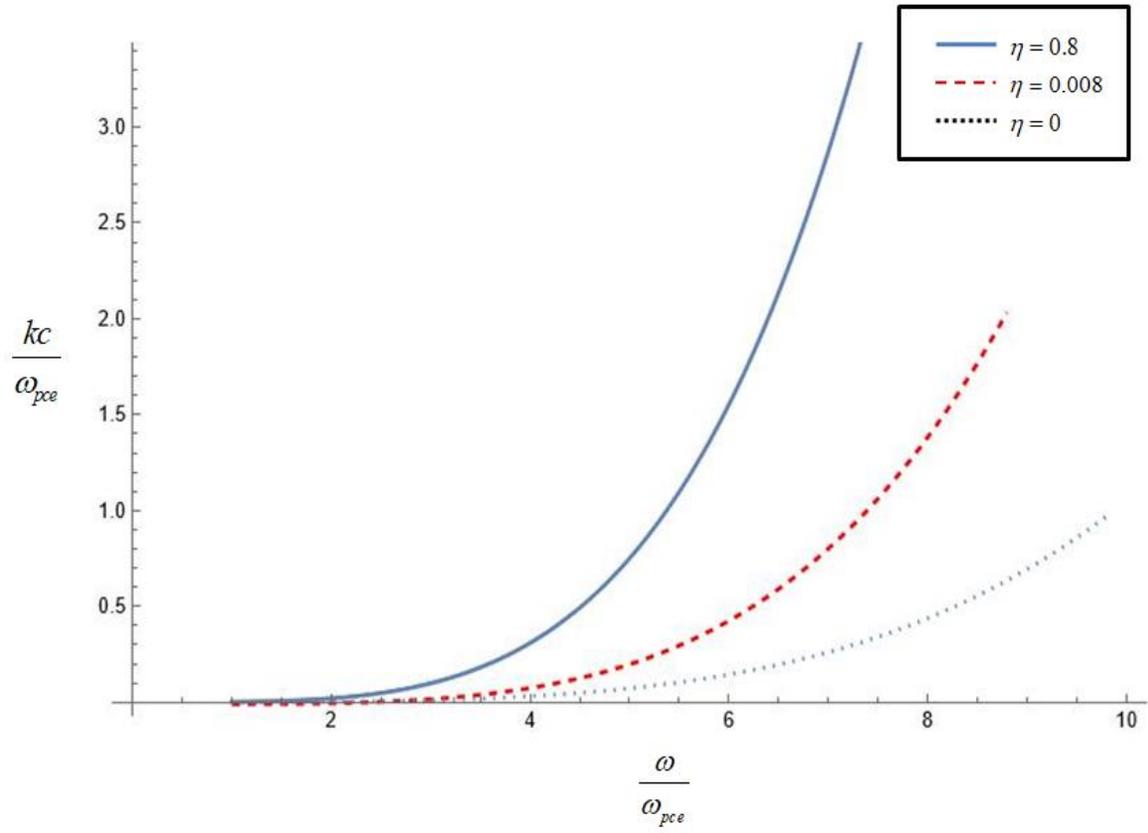

**Figure 2.**

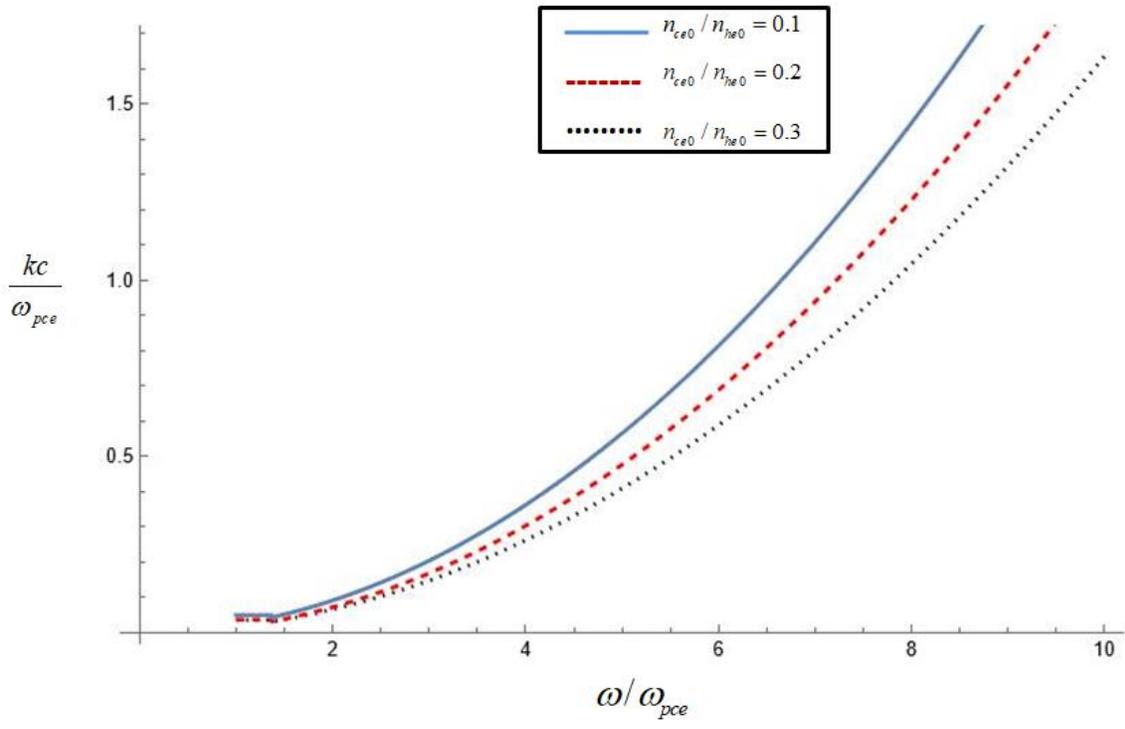

**Figure 3.**

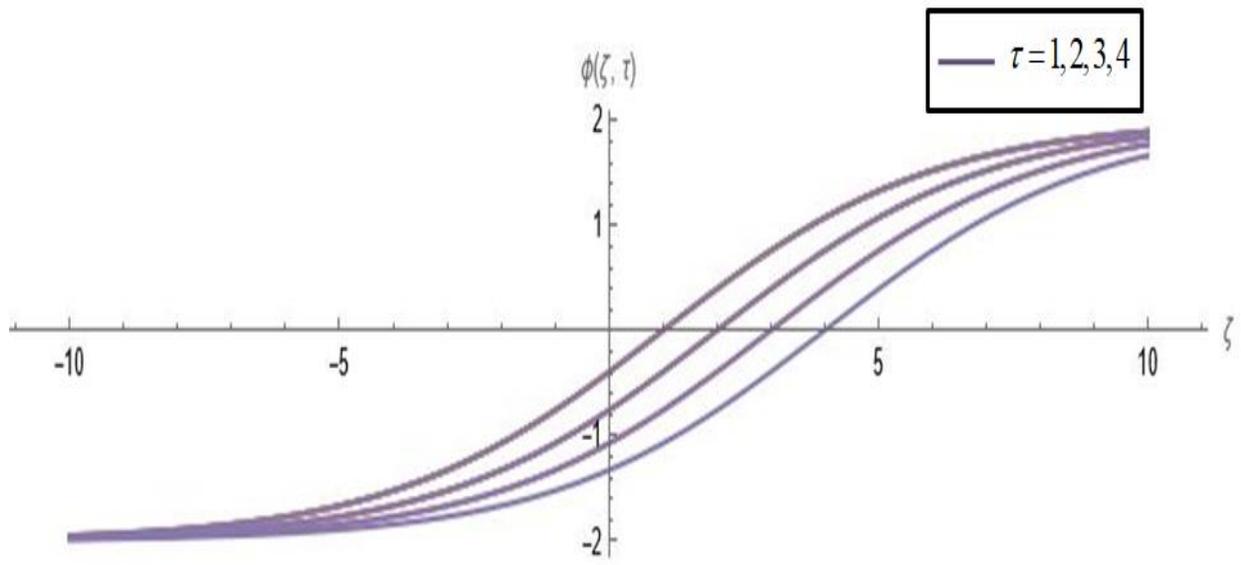

**Figure 4.**

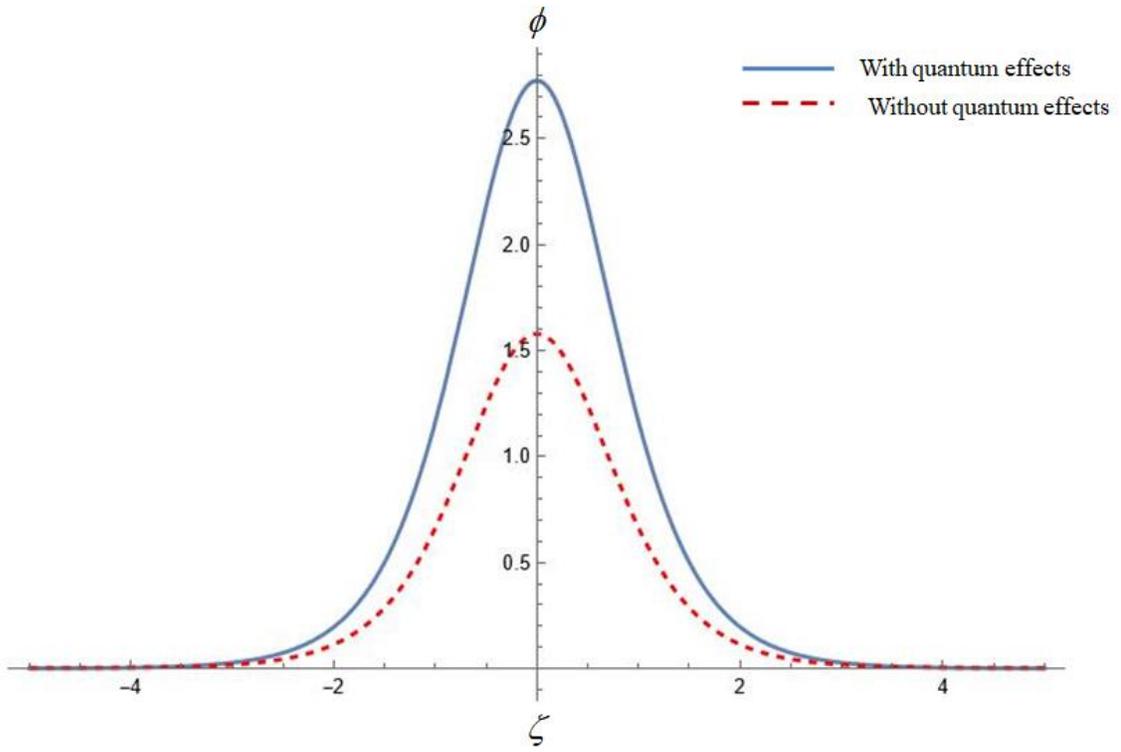

**Figure 5.**

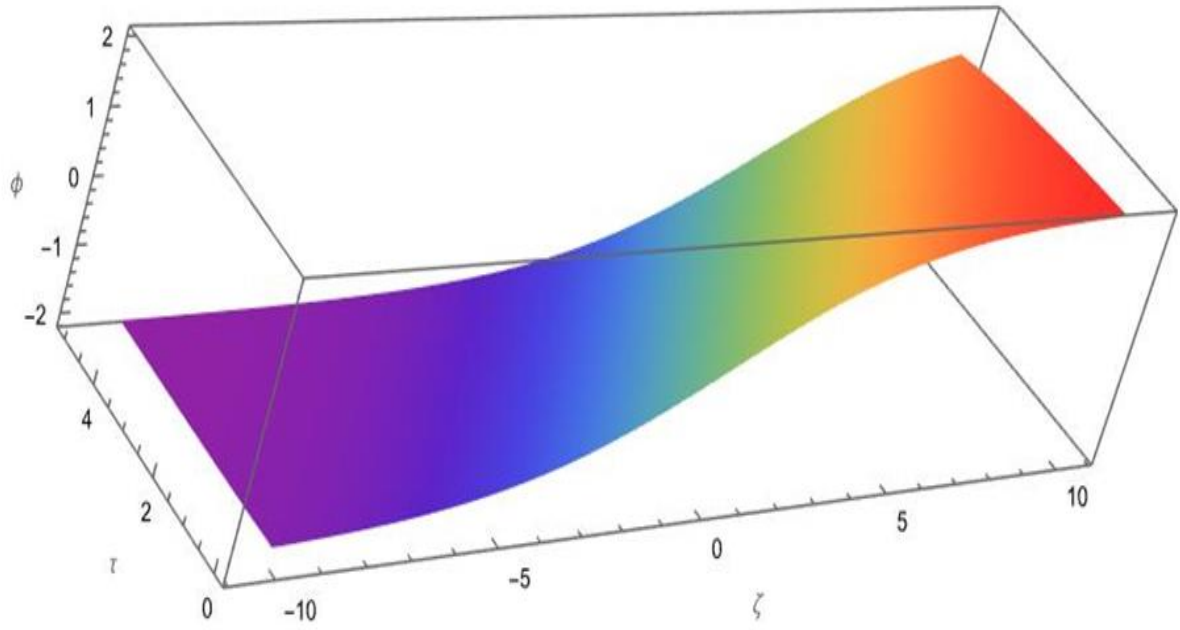

**Figure 6.**

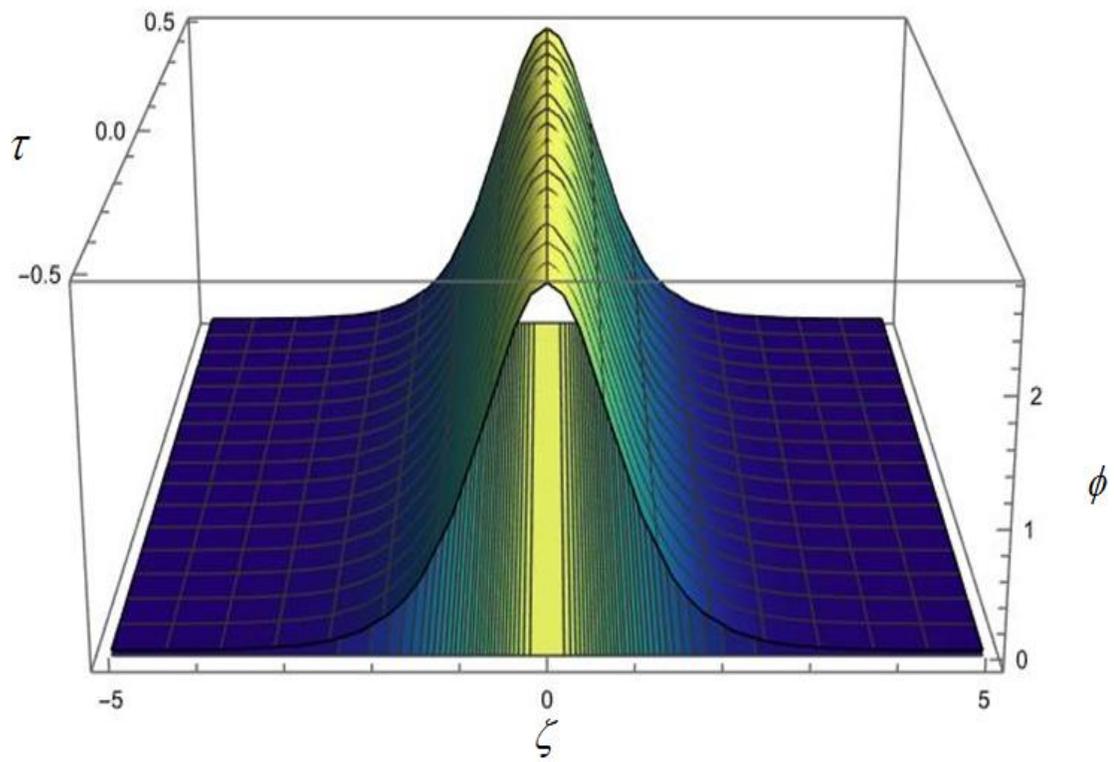

**Figure 7.**